\documentclass{article}
\usepackage{graphicx} %
\usepackage{multirow}
\usepackage{booktabs}
\usepackage{amsmath}
\usepackage{amssymb}  
\usepackage{subcaption}
\usepackage{authblk}

\usepackage{hyperref} 
\usepackage{xcolor}
\usepackage{soul}
\usepackage{transparent}
\definecolor{red}{rgb}{.95,.84,.84}
\sethlcolor{red}
\usepackage{url}
\newcommand{\sys}{S}
\newcommand{\diss}{d}
\newcommand{\learn}{\mathcal{L}} 
\newcommand{\free}{\mathcal{M}}  
\newcommand{\D}{\mathcal{D}}
\newcommand{\R}{\mathbb{R}}
\newcommand{\nsamp}{N}
\newcommand{\norm}[1]{\left\lVert#1\right\rVert}

\newcommand{\E}{\mathbb{E}}

\newtheorem{remark}{Remark}%

\title{From system models to class models: An in-context learning paradigm}
\author{Marco Forgione, Filippo Pura, Dario Piga}
\affil{IDSIA Dalle Molle Institute for Artificial Intelligence USI-SUPSI, Via la Santa 1, CH-6962 Lugano-Viganello, Switzerland.}

\date{\today}

\begin{document}

\maketitle

\noindent\rule{\textwidth}{1pt}
Please cite this version of the paper:\\ \\
M. Forgione, F. Pura and D. Piga, ``From system models to class models: An in-context learning paradigm," in \textit{IEEE Control Systems Letters,} vol. 7, pp. 3513-3518, 2023, doi: 10.1109/LCSYS.2023.3335036.
\vskip 1em \noindent
You may use the following bibtex entry:
\begin{verbatim}
@article{forgione2023from,
  author={Forgione, Marco and Pura, Filippo and Piga, Dario},
  journal={IEEE Control Systems Letters}, 
  title={From System Models to Class Models:
   An In-Context Learning Paradigm}, 
  year={2023},
  volume={7},
  number={},
  pages={3513-3518},
  doi={10.1109/LCSYS.2023.3335036}
}
\end{verbatim}
\noindent\rule{\textwidth}{1pt}

\section*{Abstract}

Is it possible to understand the intricacies of a dynamical system not solely from its input/output pattern, but also by observing the behavior of other systems within the same class? This central question drives the study presented in this paper.

In response to this query, we introduce a novel paradigm for system identification, addressing two primary tasks: one-step-ahead prediction and multi-step simulation. Unlike conventional methods, we do not directly estimate a model for the specific system. Instead, we learn a \emph{meta model} that represents a class of dynamical systems. This meta model is trained on a potentially infinite stream of synthetic data, generated by simulators 
whose settings are randomly extracted from a probability distribution. 
When provided with a   context  from a new system--specifically, an input/output sequence--the meta model implicitly discerns its dynamics, enabling predictions of its behavior.

The proposed approach harnesses the power of Transformers, renowned for their \emph{in-context learning} capabilities.  For one-step prediction, a GPT-like decoder-only architecture is utilized, whereas the  simulation problem employs an encoder-decoder structure. Initial experimental results affirmatively answer our foundational question, opening doors to fresh research avenues in system identification.

\section{Introduction}


In conventional system identification, researchers design algorithms that, given a dataset of input/output samples, return a model of the underlying data-generating mechanism.  
The typical   workflow is   closely related to supervised machine learning, with peculiarities such as the focus on dynamical systems, the choice of parsimonious representations like Linear Parameter-Varying and  hybrid  models~\cite{lovera2013lpv, mejari2020identification},  and the use of the model for complex downstream applications such as closed-loop control \cite{bombois2006least, piga2019performance}. Given the link between supervised learning and system identification, recent contributions have applied modern deep learning tools to estimate dynamical systems using neural network structures~\cite{andersson2019deep,forgione2021continuous,masti2021learning,beintema2023deep}.

The concept of \emph{meta learning}, first introduced in~\cite{schmidhuber1987evolutionary}, has gained increasing attention in the last years within the machine learning community~\cite{hospedales2021meta}. Its goal is to learn the best learning algorithm for a problem family, instead of hand-designing it with theory, intuition, or trial-and-error as done in standard machine learning. 
A way to achieve meta learning is to train architectures endowed
with \emph{in-context learning} capabilities~\cite{garg2022can, kirsch2022general, dong2022survey}.
In this framework, a few examples that serve as demonstrators are combined with a query to form the prompt, guiding the in-context learner in the generation of predictions, with the need to define neither a training algorithm  nor an inference model. 

This paper is one of the first contributions towards the
adoption of meta- and in-context learning ideas and techniques for dynamical system identification. Among the few existing works,
the Model Agnostic Meta Learning (MAML) algorithm~\cite{finn2017model}   has been recently applied in \cite{chakrabarty2023meta} for fast  model adaptation of Van der Pol oscillators with bounded parametric uncertainty from limited amount of data. 
With respect to \cite{chakrabarty2023meta}, we do not handle the problem of few-shot learning. However we consider  broader classes of dynamical systems that are not necessarily close to each other in a given parameter space representation, which is  a key assumption of MAML and its variants.  To this aim, we introduce the concept of in-context learning for system identification, where the predictions of interest are generated without deriving an explicit representation of the underlying system.  
  Although not focused on system identification, it is also worth mentioning the related work~\cite{zhan2022calibrating}, which addresses the problem of few-shot calibration of energy models for buildings by leveraging large datasets from previous calibrations on other buildings. As in our approach, an in-context learning paradigm is adopted.

The learning approach proposed in this paper can also be viewed as a \emph{meta-modeling} framework, where a model describing an entire class of  systems, rather than a particular element of the class, is learned. 
The behaviour of the actual data-generating system is  inferred from a \emph{context} of input/output data, and thus  downstream tasks like one-step-ahead prediction or simulation are solved, which would otherwise require to learn a traditional system-specific model for each dataset.   
Our guess is that training from  a large amount of dynamical systems allows  the meta model to implicitly learn  latent (not easily interpretable) features relevant for entire classes of dynamical systems.  

Training of the meta model is performed on data generated by different, but related systems. We assume to have access to an \emph{infinite stream} of dynamical systems generating input/output  datasets. In practice,  data may come from  simulators. 
We can therefore generate arbitrary amount of synthetic examples, by varying software settings representing plausible scenarios (e.g., physical parameters and disturbances) according to prior knowledge and insights. From this stream of data, we can learn to make optimal predictions for the considered  class without modeling each individual system. The trained meta model can then be applied to make predictions on a \emph{particular} dataset collected from a real system.
 
To guarantee substantial representational power of the meta model, we leverage Transformer architectures analogous to those employed in Natural Language Processing \cite{vaswani2017attention,radford2019language}. In this paper, Transformers are specialized to handle real-valued input/output sequences and address two common  system identification tasks: one-step-ahead prediction and multi-step-ahead simulation, using distinct architectures for the two problems. For one-step-ahead prediction, we employ a GPT-like encoder-only structure~\cite{radford2019language}, which presently stands as the state-of-the-art for 
text generation.   Conversely, for multi-step-ahead simulation, we employ an encoder-decoder Transformer, drawing inspiration from the  architecture introduced in~\cite{vaswani2017attention}, which stands as the current benchmark for language translation. In our work, the encoder output can be interpreted as an implicit representation of the  system, enabling the decoder to simulate the system’s response to new input sequences.

As we finalized this manuscript, we came across a recent independent work~\cite{balim2023can} that tackles the  one-step-ahead prediction problem using an in-context learning strategy closely mirroring ours with a GPT network. For multi-step-ahead simulation, we are not aware of any contribution that addresses the problem within an in-context learning framework or that uses the encoder-decoder Transformer architecture presented in this paper.

To ensure the replicability and reproducibility of our research results, and to encourage further contributions to the field, we have made the PyTorch implementation of all methodologies and results available in the GitHub repository \href{https://github.com/forgi86/sysid-transformers}{https://github.com/forgi86/sysid-transformers}.

\section{Learning framework}


In traditional  system identification, we are given a dataset $\D_{\rm train} = (u_{1:\nsamp}, y_{1:\nsamp})$ 
generated by a \emph{fixed} unknown dynamical system $\sys$, with $u_k \in \mathbb{R}^{n_u}$ (resp. $y_k  \in \mathbb{R}^{n_y}$) representing the system's input (resp. output) at time step $k$. The objective is to estimate a model $M$ of $\sys$ from the dataset $\D$ and available prior assumptions on the system $\sys$, typically formalized in terms of a parametric model structure $\{M(\theta), \theta \in \Theta \subseteq \mathbb{R}^{n_\theta} \}$.
The model is chosen by minimizing a cost function over the training data
\begin{equation}
\label{eq:sysid_objective}
\theta^* = \learn({\D_{\rm train}}) = \arg \min_{\theta \in \Theta} \diss(\D_{\rm train}, M(\theta)),
\end{equation}
where $\diss$ is a measure of dissimilarity between the measured data and the output's predictions of the model, such as a one-step or multi-step simulation loss. Note that in \eqref{eq:sysid_objective} we emphasized the fact that the optimal model parameter $\theta^*$ is a function $ \learn$ of the training dataset $\D_{\rm train}$.

For the sake of concreteness, $M(\theta)$ could be the 
state-space model:
\begin{subequations}
\label{eq:ss}
\begin{align}
x_{k + 1} &= f_\theta(x_k, u_k)\\
\hat y_k &= g_\theta(x_k),
\end{align}
\end{subequations}
where $x_k \in \R^{n_x}$ is a hidden state variable and $f_{\theta}$, $g_{\theta}$ are functions parametrized by  $\theta$ (e.g., neural networks). The cost function quantifying model fitness could be 
the simulation error loss:
\begin{equation}
\diss(\D_{\rm train}, M(\theta)) = \frac{1}{\nsamp} \sum_{i=1}^{\nsamp} \norm{y_i- \hat y_{i}}^2, 
\end{equation}
where $\hat y_{i}$ is obtained by iterating the model's equations~\eqref{eq:ss} up to time step $i$, using the input sequence $u_{1:i-1}$.
Model performance may be assessed by evaluating the dissimilarity metrics $\diss$
on a distinct validation dataset $\D_{\rm val}$:
$\diss(\D_{\rm val}, M(\theta^*))$, which  gives an   estimate of the model's \emph{generalization error}.

In the learning framework considered in this paper and discussed in the following sections, we have a prior distribution for dynamical systems, which is used to generate a sequence of such systems. We also have a prior distribution for input signals that produces an input time sequence, exciting the generated dynamical systems and thus resulting in a set of input-output trajectories. This allows us to generate a potentially infinite number of input-output training datasets. 

Two learning frameworks are discussed in the following. First, in Section~\ref{sec:model_based} we introduce the concept of model-based meta learning that directly produces as an output the optimized parameters of a model from  given system's input/output sequence. Then, in Section~\ref{sec:model_free_model_based} we discuss the model-free in-context learning approach, which represents the main contribution of our work. 

\subsection{Model-based meta learning for system identification} \label{sec:model_based}
In the setting considered in  this 
paper, we assume to have access to an \emph{infinite stream} of input/output pairs $\{\D^{(i)} = (u_{1:N}^{(i)}, y_{1:N}^{(i)}), i=1,2,\dots, \infty \}$, each obtained by processing a randomly generated input signal $u_{1:N}^{(i)}$ through a randomly instantiated dynamical system
$\sys^{(i)}$, possibly corrupted by a likewise random disturbance. In other words, we can sample (possibly synthetic) datasets $\D^{(i)}$ from an underlying  distribution $p(\D)$.

Since the datasets $\D^{(i)}$ are related with one another,
partial knowledge transfer from one dataset to the other is possible, and one could then exploit it to  improve the identification performance as more datasets are 
observed. Having access
to an infinite data stream, we should be able at the limit to \emph{learn a learning algorithm} that
is optimal (w.r.t. a given criterion, e.g. one-step prediction)  in some probabilistic sense (e.g., expected value) for the dataset distribution $p(\D)$.

Let us introduce a parametrized \emph{family of 
learning algorithms} $\{\learn_\phi(\cdot), \phi \in \Phi\}$. The learning algorithm $\learn_{\phi}(\cdot)$ is a map 
from a training dataset to a dynamical model.   
We seek then the learning rule that optimizes the fitting objective in an expected value sense over the dataset distribution $p(\D)$:
\begin{equation}
	\label{eq:meta_objective_exp_noval}
    \phi^{*} = \arg \min_\phi \E_{p(\D)} \left [\diss\left ( \D, \learn_\phi(\D) \right ) \right ].
\end{equation}

The function $\learn_\phi$ could be either an explicit map from datasets to model parameters, e.g., a neural network  from datasets to model parameters with tunable weights $\phi$~\cite{gordon2018metalearning},
or an \emph{inner} optimization algorithm with tuning knobs $\phi$ to be applied to the dataset to obtain model parameters.  
A well-known example of the second approach is  MAML \cite{finn2017model}, where $\phi$ represents the initial condition for an inner gradient-based optimization carried out on training datasets.

The \emph{nested} nature of meta learning can be seen from the formulation in~\eqref{eq:meta_objective_exp_noval}, where the outer objective
to be minimized w.r.t $\phi$ is the performance index of an the inner learning algorithm $\learn_\phi$, that in turns is designed to optimize a performance index over the training dataset $\D$. 

Actually, the objective \eqref{eq:meta_objective_exp_noval}
needs to be modified to avoid the possibility of learning 
an overfitting algorithm, which 
simply memorizes each training dataset.
To this end, according to the meta-learning practice~\cite{hospedales2021meta},  we can split each dataset $\D^{}$ into training and validation 
portions $\D^{}_{\rm train}$ and $\D^{}_{\rm val}$, respectively, and optimize $\phi$ according to the 
criterion:
\begin{equation}
	\label{eq:meta_objective_exp}
    \phi^{*} = \arg \min_\phi 
  \E_{p(\D)} \left [ \diss\left (\D_{\rm val}, \learn_\phi(\D_{\rm train})\right ) \right],
\end{equation}
where learning and performance evaluation of the algorithm $\learn_\phi$ are executed on distinct splits of the dataset $\D$.

As known,  since the expectation in~\eqref{eq:meta_objective_exp} is intractable (the dataset distribution $p(\D)$ may be very complex, and it may only available through samples anyway), the  expectation is approximated
through the sample average:
\begin{equation}
	\label{eq:meta_objective_mc}
  \E_{p(\D)} \left [ \diss\left (\D_{\rm val}, \learn_\phi(\D_{\rm train})\right ) \right]  \approx  \frac{1}{b} \sum_{i=1}^b 
 \diss\left (\D^{(i)}_{\rm val}, \learn_\phi(\D^{(i)}_{\rm train})\right ), 
\end{equation}
where $b$ is the sample  size, and  the datasets  $\D^{(i)}$ are drawn from the probability distribution $p(\D)$.

Having access to an infinite stream of datasets $\D^{(i)}$, we can actually follow a ``pure'' \emph{stochastic gradient descent} approach and approximate the expected value in \eqref{eq:meta_objective_exp} with a sample average on $b$ datasets resampled \emph{at each iteration}. This is in contrast with standard supervised learning, where \emph{gradient descent} relies on a finite number of training instances. 

Having access to an infinite stream of datasets rules out the risk of \emph{meta-overfitting}, namely of choosing a learning rule $\learn_\phi$ too much tailored to a specific 
subset of datasets that does not generalize to the whole distribution.
 
\begin{remark}
Overfitting and meta-overfitting are separate concepts corresponding to different risks. The former is related to learning a too complex $\learn_\phi$ that tends to memorize specific properties of each dataset $\D^{(i)}_{\rm train}$, without modeling the underlying mechanism $\sys^{(i)}$ and thus failing to generalize well to further data $\D^{(i)}_{\rm val}$ from the same systems.
The latter is related to learning a rule $\learn_\phi$ too tailored to the particular datasets $\D^{(i)}$ seen in meta-training, which is less effective on other datasets from the same distribution $p(\D)$. 
Plain overfitting  may be dealt with a train-validation split of the meta objective as done in \eqref{eq:meta_objective_exp}, while meta overfitting is circumvented by training in a pure stochastic gradient descent setting, with different datasets $\D^{(i)}$ sampled at each iteration. 
\end{remark}

\subsection{Model-free in-context learning for system identification}
\label{sec:model_free_model_based}
In model-based meta learning, the algorithm $\learn_\phi$ is a map taking the training dataset $\D^{(i)}_{\rm train}$ as an input and returning  a model $M^{(i)}$ describing the behaviour of the dynamical  system $\sys^{(i)}$. 

In the model-free in-context  learning approach proposed in this paper, we learn instead a map $\free_{\phi}$ (called meta model) which processes  portions of the dataset $\D^{(i)}$ and directly reproduces the outputs  of interest, without generating an intermediate, explicit representation (namely, model) of the systems. 

The two instances that are discussed and experimentally validated in this paper consist of:
\begin{itemize}
    \item \textbf{model-free one-step-ahead prediction:}  In this problem, for each input/output sequence of a dataset $\D^{(i)}$ and for each time step $k$, the meta-model $\free_{\phi}$ digests partial input/output pairs ($u_{1:k}^{(i)}, y_{1:k}^{(i)}$) up to time $k$ and produces predictions $\hat y_{k+1}^{(i)}$ for the output at time step $k+1$:
\begin{equation}
\label{eq:meta_one_step}
\hat y_{k+1}^{(i)} = \free_\phi(u_{1:k}^{(i)}, y_{1:k}^{(i)}).
\end{equation}
\item \textbf{model-free simulation:} In this problem, the model model $\free_{\phi}$ receives the input/output  ($u_{1:m}^{(i)}, y_{1:m}^{(i)}$) up to time step $m$
and a test input sequence (\emph{query})  $u_{m+1:N}^{(i)}$ from time $m+1$ to $N$ and produces the corresponding output sequence 
$\hat y_{m+1:N}^{(i)}$. 
\begin{equation}
\label{eq:model_free_sim}
\hat y_{m:N}^{(i)} = \free_\phi(u_{1:m-1}^{(i)}, y_{1:m-1}^{(i)}, u_{m:N}^{(i)}).
\end{equation}
We remark that in a model-based setting, this problem would be tackled by learning a system-specific model from the input/output pair ($u_{1:m}^{(i)}, y_{1:m}^{(i)}$), and then applying this model in simulation mode on $u_{m+1:N}^{(i)}$. Furthermore, note that $\free_\phi$ in \eqref{eq:model_free_sim} may be seen as a simulation model with structure:
$M^{(i)}(\cdot) 
= \free_\phi(u_{1:m}^{(i)}, y_{1:m}^{(i)}, \cdot)$.
\end{itemize}

 To solve the model-free one-step-ahead prediction and simulation problems mentioned above, the meta-model $ \free_\phi$ is expected to ``understand'' (to a certain degree) the data generating mechanism $\sys^{(i)}$ from the provided \emph{context} ($u_{1:k}^{(i)}, y_{1:k}^{(i)}$) (respectively, ($u_{1:m}^{(i)}, y_{1:m}^{(i)}$)), but it does not return an explicit representation in a model form. Rather, as discussed in the  Section~\ref{sec:train}, $\free_\phi$  is trained to output the one-step-ahead predictions (respectively, the output sequence continuation) of interest directly.

 In the rest of the paper  we focus  on model-free in-context learning, which represents our core contribution and the novel system identification paradigm proposed in this work. 

\section{Model-free in-context learning: architectures and training}
\label{sec:train}

Unlike model-based meta learning, model-free in-context learning is conceptually closer  
to standard supervised learning where $\free_\phi$ is  a model directly mapping from features to targets rather than a learning algorithm. %
Training can be seen as a single-level optimization, rather than a nested learning procedure as in 	\eqref{eq:meta_objective_exp_noval}.
However, when faced with system variability in the dataset distribution, the problem setting still necessitates that $\free_\phi$ be as powerful as a model learning algorithm. Therefore, in this paper, we employ Transformer architectures, which are currently state-of-the-art for in-context learning, especially in NLP applications. While other neural network architectures suitable for time sequence processing can also be utilized (like LSTM), preliminary   experiments (not reported in this document) have shown that  Transformer architectures consistently achieve better performance.

\subsection{Decoder-only Transformer for one-step-ahead prediction} \label{sec:train_onestep}

\begin{figure}[!bt]
\includegraphics[width=.8\textwidth]{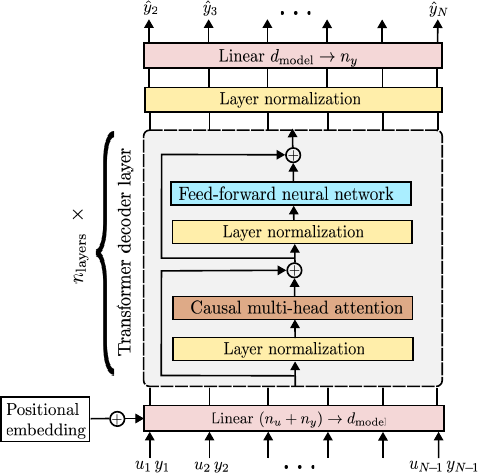}
\caption{GPT-like decoder-only Transformer   for one-step-ahead prediction. Differences w.r.t. plain GPT-2 for text generation~\cite{radford2019language, nanoGPT} are highlighted in \hl{pink}.}
\label{fig:decoder_only_arch}
\end{figure}

 A  decoder-only Transformer architecture derived from GPT-2~\cite{radford2019language} is developed.
 The model structure is fully specified by the choice of the hyper-parameters characterizing a Transformer, namely:  number of layers $n_{\rm layers}$; number of units in each layer $d_{\rm model}$; number of heads $n_{\rm heads}$; and the context window length $n_{\rm ctx}$. 
The standard Transformer   is modified to process the real-valued input/output sequences generated by dynamical systems, instead of the sequence of symbols (word tokens) needed for natural language modeling. To this aim, with respect to plain GPT-2, the initial token embedding layer is replaced by a linear layer with $n_u+n_y$ inputs and $d_{\rm model}$ outputs, while the final layer is replaced by a linear layer with $d_{\rm model}$ inputs and $n_y$ outputs. 
The overall architecture is visualized in Fig.~\ref{fig:decoder_only_arch}.

By considering a quadratic loss on the output, the  weights $\phi$ of the meta-model $\free_{\phi}$ are obtained by minimizing over $\phi$ the loss:
\begin{equation}
	\label{eq:one_step_objective}
     \E_{p(\D)} 
    \left [ \sum_{k=1}^{\nsamp-1}
    \norm{y_{k+1} -
    \free_\phi (y_{1:k}, u_{1:k})
    }^2
    \right ],
\end{equation}
where the expected value is approximated with $b$ samples as
\begin{equation}
	\label{eq:eq:one_step_objective_samples}
\E_{p(\D)} 
    \left [ \sum_{k=1}^{\nsamp-1}
    \norm{y_{k+1} -
    \free_\phi (y_{1:k}, u_{1:k})
    }^2
    \right ]
    \approx \frac{1}{b}
    \sum_{i=1}^b
    \sum_{k=1}^{\nsamp-1}
    \norm{y_{k+1}^{(i)} -
    \free_\phi (y^{(i)}_{1:k}, u_{1:k}^{(i)})}^2
    .
\end{equation}

Note that the input of the meta-model $\free_\phi$ encompasses the entire input/output sequence from time $1$ up to time $N-1$, as illustrated in Fig. \ref{fig:decoder_only_arch}. However, due to the causal multi-head attention layer, the output at time $k+1$ depends only on the past input/output samples up to time $k$. For this reason, albeit with some abuse of notation, only the sequence $u_{1:k}, y_{1:k}$ appears as the input of $\free_\phi$ in eqs. \eqref{eq:one_step_objective} and \eqref{eq:eq:one_step_objective_samples}.

\subsection{Encoder-Decoder Transformer for simulation}
\label{sec:train_sim}
\begin{figure}[!bt]
\includegraphics[width=.99\textwidth]{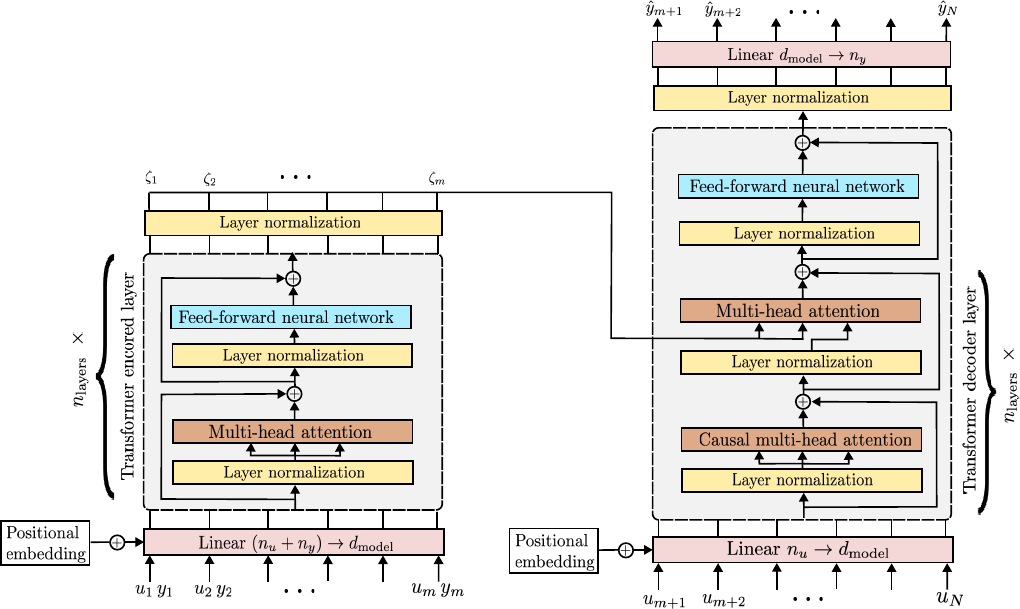}
\caption{Encoder-decoder Transformer  for multi-step-ahead simulation. Main differences w.r.t. the standard Transformer architecture for language translation~\cite{vaswani2017attention} are highlighted in \hl{pink}.}
\label{fig:encoder_decoder_arch}
\end{figure}

An encoder-decoder Transformer similar to the one originally introduced for language translation in~\cite{vaswani2017attention} and adapted to the model-free in-context simulation task is used. The overall architecture is visualized in Fig. \ref{fig:encoder_decoder_arch} and consists in: ($i$) an encoder that processes 
$u_{1:m}, y_{1:m}$ (without causality restriction) and generates an embedding sequence $\zeta_{1:{m}}$; ($ii$) a decoder  that processes $\zeta_{1:{m}}$ and test input  $u_{m+1:\nsamp}$ (the latter with causal restriction) to produce the 
sequence of predictions $\hat y_{m+1:\nsamp}$.
Similarly to the one-step-ahead prediction task discussed in Section \ref{sec:train_onestep}, the standard encoder-decoder Transformer is modified to process  real-valued input/output sequences.

In a model-based interpretation, the output of the encoder $\zeta_{1:{m}}$ may be seen as a hidden  representation of the system $\sys^{(i)}$ that is used as an implicit ``model parameter'' enabling the decoder to simulate the system's response to the sequence $u_{m+1:\nsamp}$. 

Similarly to  one-step-ahead prediction case, the  weights $\phi$ of the meta-model $\free_{\phi}$ are obtained by minimizing over $\phi$  the loss 
\begin{equation}
	\label{eq:simulation_objective}
     \E_{p(\D)} 
    \left [ 
    \norm{y_{m+1:\nsamp} -  \free_\phi (u_{1:m}, y_{1:m}, u_{m+1:\nsamp})
    }^2
    \right ].
\end{equation}
As in~\eqref{eq:one_step_objective}, a sample-based approximation over systems $\sys^{(i)}$ 
and  datasets $\D^{(i)}$  is used to approximate the expected value~\eqref{eq:simulation_objective}:
\begin{multline}
	\label{eq:simulation_objective_samples}
     \E_{p(\D)} 
    \left [ 
    \norm{y_{m+1:\nsamp} -  \free_\phi (u_{1:m}, y_{1:m}, u_{m+1:\nsamp})
    }^2
    \right ]
    \approx \\
    \frac{1}{b}
    \sum_{i=1}^b
    \norm{y_{m+1:\nsamp}^{(i)} - \free_\phi (u_{1:m}^{(i)}, y_{1:m}^{(i)}, u_{m+1:\nsamp}^{(i)})}^2
    .
\end{multline}

\section{Examples}
\label{sec:examples}

In this section, we present results on  one-step-ahead prediction and multi-step-ahead simulation for two classes of dynamical systems, namely
Linear Time Invariant (LTI) and Wiener-Hammerstein (WH). The latter represents a block-oriented description of several real-world nonlinear dynamical systems~\cite{giri2010block}.  
The architectures are trained by minimizing the loss functions 
\eqref{eq:one_step_objective} and \eqref{eq:simulation_objective} for one-step-ahead prediction and multi-step-ahead simulation, respectively.
For numerical optimization, the \emph{AdamW} algorithm~\cite{loshchilov2017decoupled} is employed. 

The software has been developed in the PyTorch deep learning framework and it is fully available in the GitHub repository \url{https://github.com/forgi86/sysid-neural-transformers}. The code of the decoder-only architecture is adapted from the  GPT-2 implementation by A.~Karpathy~\cite{nanoGPT}, while we developed the code of the encoder-decoder architecture mostly from scratch.

Computations are performed on a server of the IDSIA laboratory equipped with 2 64-core AMD EPYC 7742 Processors, 256 GB of RAM, and 4 Nvidia RTX 3090 GPUs. In all the experiments, the utilized resources have been limited to 10 CPU threads and 1 GPU.

\paragraph{Linear Time Invariant system class (LTI)}
We consider stable single-input-single-output linear time invariant dynamical systems in state-space form, with random order between 1 and 10. The state-space matrices $A$,$B$,$C$,$D$ are randomly generated with the constraint that the eigenvalues of matrix $A$ lie in the region within the complex unit circle with magnitude in the range $(0.5,0.97)$ and phase in the range $(-\pi/2 , \pi/2)$. 

\paragraph{Wiener-Hammerstein system class (WH)}
We consider stable Wiener-Hammerstein dynamical systems~\cite{giri2010block} with structure $G_1$--$F$--$G_2$, where $G_1$, $G_2$ are SISO LTI blocks and $F$ is a static non-linearity. $G_1$ and $G_2$ are randomly generated with order comprised between 1 to 5, and with the same magnitude/phase constraint on the eigenvalues used for the LTI class introduced in the previous paragraph. The static non-linearity $F$ is defined as a feed-forward neural network with one hidden layer and random parameters generated from a  Gaussian distribution with  Kaiming scaling. 
\vskip 1em
For both system classes, the input signal applied in our experiments has a white Gaussian distribution with zero mean and unit variance. Each dataset $\D^{(i)}$ is thus constructed by sampling a random input sequence  $u_{1:\nsamp}^{(i)}$ and by applying it to a randomly sampled system $
\sys^{(i)}$ (either LTI or WH), thus obtaining the output sequence 
$y_{1:\nsamp}^{(i)}$. Finally, for easier numerical optimization, the system output $y_{1:\nsamp}^{(i)}$ is scaled to have zero mean and unit variance.

\subsection{One-step-ahead prediction}
For one-step-ahead prediction, we applied the decoder-only Transformer architecture
described in Section~\ref{sec:train_onestep}, with different choices of the hyper-parameters.
The results are summarized in Table~\ref{tab:one_step}, where we report the hyper-parameters of the meta model and of the optimization, the train time, and the achieved one-step-ahead \emph{root mean square error} ($\rm rmse$).

\begin{figure}
    \centering

    \begin{subfigure}[t]{0.99\textwidth}
        \centering
        \includegraphics[width=\textwidth]{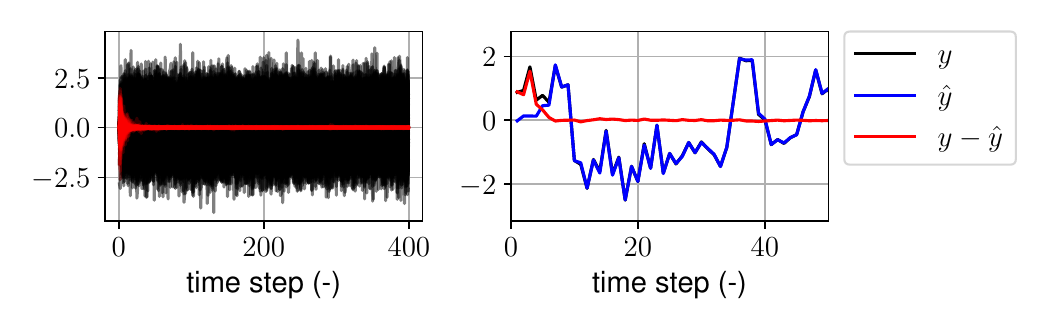}
        \caption{LTI class: prediction results on 256 randomly sampled systems {superposed} (left) and on a particular system (right).}
        \label{fig:lin_sim_batch_single}
    \end{subfigure}
    \hfill
    \begin{subfigure}[t]{0.99\textwidth}
        \centering
        \includegraphics[width=\textwidth]{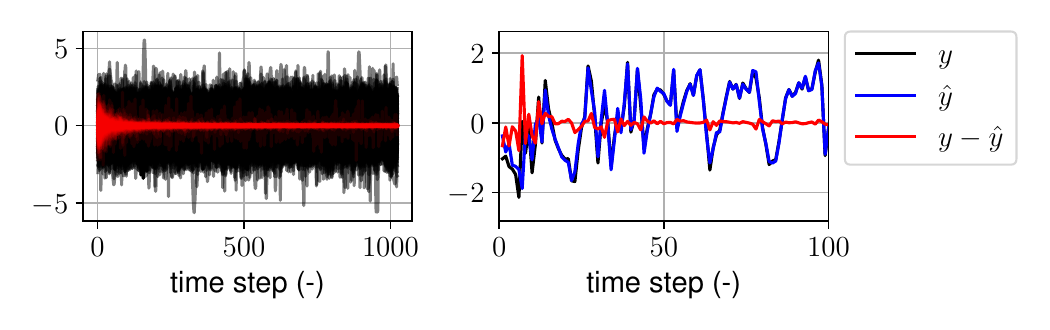}
        \caption{WH class: prediction results on 256 randomly sampled systems {superposed} (left) and on a particular system (right).}
        \label{fig:wh_sim_batch_single}
    \end{subfigure}
    
    \caption{One-step-ahead prediction on the LTI (top row) and WH (bottom row) model classes. Actual output $y$ (black), simulated output $\hat y$ (blue), and simulation error $y-\hat y$ (red). }
    \label{fig:one_step_prediction}
\end{figure}

As for the LTI system class, we train a single architecture with 1.68 million parameters.
Gradient-based optimization is performed over 300'000 iterations, which required about 2 hours on our server.
The performance of the trained model is highlighted in Figure~\ref{fig:one_step_prediction} (top row). The left plot denotes the prediction error $y - \hat y$ achieved by the trained Transformer over 256 randomly extracted LTI systems.
It is evident that the prediction error decreases for increasing time step $k$, as the context 
 $(u_{1:k-1}, y_{1:k-1})$ available to make the prediction  $\hat y_k$   becomes larger.
Around time step 20, the error's magnitude has decreased to a very small value for all the 256 visualized cases. In the right plot, we show in more detail the trajectory of a particular system, reporting the true output $y$, the predicted output $\hat y$, and the prediction error $y - \hat y$ altogether. It appears that, for this particular system, the Transformer is able to make nearly-optimal predictions within about 10 time steps.

As for the WH system class, we tested a medium- and a large-size Transformer with total number of parameters  2.44 million  and 85.74 million, respectively.
Figure~\ref{fig:one_step_prediction} (bottom row) visualizes the performance of the large-size Transformer over 256 randomly extracted
WH systems.
Similarly to the LTI case, we report ($i$) the prediction error $y - \hat y$ of the 256 system realizations in the left subplot and $(ii)$ the true output $y$, the predicted output $\hat y$, together with the prediction error 
$y - \hat y$ of one particular realization in the right subplot. We observe the same qualitative behavior previously seen in the LTI case, with the 
prediction error generally decreasing for increasing step index. 
With respect to the LTI case, the error magnitude decreases more slowly and requires about 200 steps to stabilize. This result is reasonable since  the complexity of the WH class is much higher than the LTI one. A richer context of data is then needed to discern a particular system within the WH class, and thus to be able to make good predictions.

\begin{table*}
    \centering
    \caption{One-step-ahead prediction: settings and outcomes. The  reported  $\rm {rmse}$ refers to one-step-ahead prediction.}
    \label{tab:one_step}
    \resizebox{0.9\textwidth}{!}{%
    \begin{tabular}{lccccccccc}
        \toprule
        $p(\D)$ & $n_{\text{param}}$ & $n_{\text{layers}}$ & $n_{\text{heads}}$ & $d_{\text{model}}$ & $n_{\text{ctx}}$ & $n_{\rm itr}$ & batch size $b$ & train time & rmse \\
        \midrule
        LTI & 1.68~M & 4 & 4 & 128 & 400 & 300K & 32 & 2~h & 0.009 \\
        WH & 2.44~M & 12 & 4 & 128 & 600 & 1M & 32 & 0.8~d & 0.041 \\
        WH & 85.74~M & 12 & 12 &768 & 1024 & 10M & 20 & 7.3~d & 0.018 \\
        \bottomrule
    \end{tabular}%
    }
\end{table*}

\subsection{Simulation}
\label{sec:exp_sim}
For the simulation task, we applied the encoder-decoder Transformer architecture described in Section~\ref{sec:train_sim}.  We generate sequences of length $\nsamp=500$, use the first $m=n_{\rm ctx, enc} = 400$ samples for the 
encoder, and simulate over the last $\nsamp-m=n_{\rm ctx, dec}=100$ samples in the decoder.

Results are summarized in Table~\ref{tab:sim}, where we report the hyper-parameters of the meta model and of the optimization, the train time, and the simulation \emph{root mean square error} (rmse). 
It is worth mentioning that, to train the meta model on the WH class, we {warm-started} the optimization with the previously-trained weights of the LTI class. 

\begin{figure}
    \centering
    \begin{subfigure}[t]{0.99\textwidth}
        \centering
        \includegraphics[width=\textwidth]{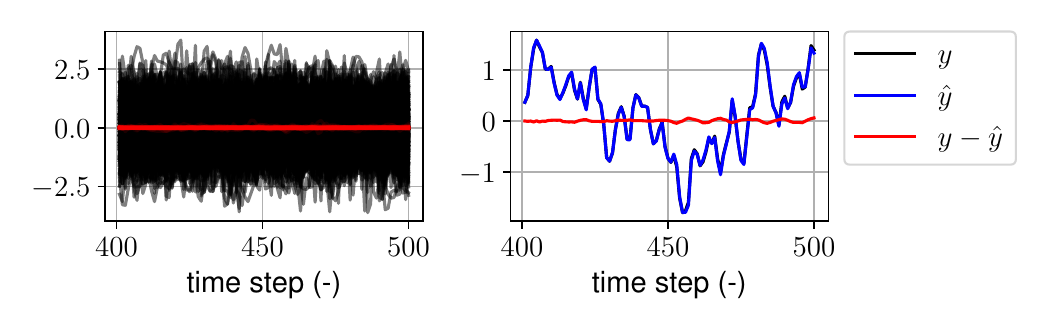}
        \caption{LTI class: simulation results on 256 randomly sampled systems {superposed} (left) and on a particular system (right).}
        \label{fig:lin_sim_batch}
    \end{subfigure}
    \hfill
    \begin{subfigure}[t]{0.99\textwidth}
        \centering
        \includegraphics[width=\textwidth]{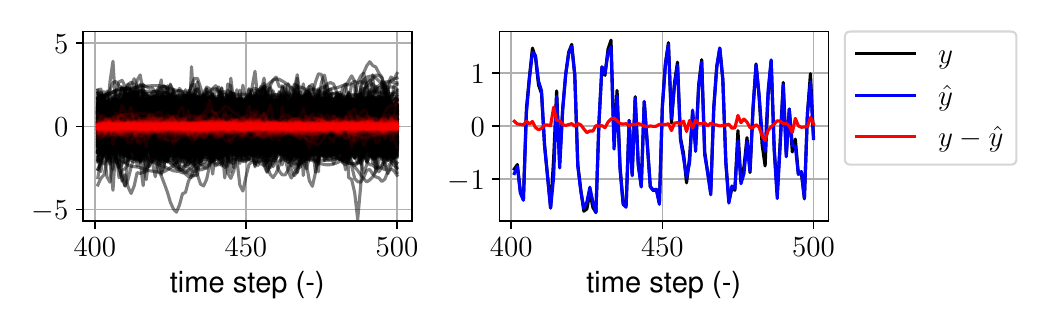}
        \caption{WH class: simulation results on 256 randomly sampled systems superposed (left) and on a particular system (right).}
        \label{fig:lin_sim_single}
    \end{subfigure}
    
    \caption{Simulation on the LTI (top row) and WH (bottom row) model classes. Actual output $y$ (black), simulated output $\hat y$ (blue), and simulation error $y-\hat y$ (red). }
    \label{fig:sim}
\end{figure}

The performance of the trained meta models (both for the LTI and WH classes) is visualized in Figure~\ref{fig:sim}. The left panels show the simulation error $y - \hat y$ achieved by the trained Transformers over $256$ randomly extracted LTI (top row) and WH (bottom row) systems.
Unlike one-step-ahead prediction, the error does not decrease for increasing simulation time step. Indeed, 
 the context $(u_{1:m}, y_{1:m})$ fed into the encoder has already being processed into the encoder embedding $\zeta_{1:m}$ before providing the   input-only query sequence $u_{m+1:N}$, which does not contain further information on the system.

{For the LTI class, we also tested the performance under a \emph{distribution shift}, where the eigenvalues of matrix $A$ are extended to the region in the complex plane with amplitude $(0.2, 0.99)$ and phase ($-3/4 \pi, 3/4 \pi)$. The average $\rm rmse$ achieved by the Transformer trained in the nominal conditions increases to $0.066$, vs. $0.009$ of the nominal case.}

{
Finally, we compare the simulation performance of the encoder-decoder transformer with a traditional system identification approach where we fit system-specific models from scratch on the first $n_{\rm ctx, enc}=400$ samples, and then use these models to simulate the output on the following $n_{\rm ctx, dec}=100$ samples. For the LTI class, we apply subspace identification~\cite{van2012subspace}, as implemented in the \texttt{n4sid} MATLAB function. For the WH class, we assume perfect knowledge of the architecture (order of the linear blocks and structure of the static non-linearity) and fit WH models with a non-linear least squares (NLSQ) approach. Specifically, we minimize the mean squared simulation error using the Adam optimizer and exploiting the approach in~\cite{forgione2021dynonet} for fast differentiation of the linear blocks. 
In this analysis, we also consider the effect of a white Gaussian noise term corrupting the measured output $y_{1:m}$ in the context.
We then repeat the identification procedure for increasing values of the noise standard deviation $\sigma_e$, for all the 256 systems considered in Fig.~\ref{fig:lin_sim_batch_single}, \ref{fig:wh_sim_batch_single}. 
Moreover, we assess the performance of the Transformer (which is trained on noise-free data) fed with the same noisy datasets.
Results are summarized in Table~\ref{tab:noise_effect}.
It is worth noting that (i) the traditional system identification approach approach exploits knowledge of the WH system structure, while we only need to draw samples from the dataset distribution to train our Transformer and ($ii$) fitting each WH model from scratch takes approximately 60 seconds, while we  obtain { simulations for} all the 256 WH datasets simultaneously in just 2 seconds with the Transformer, without the need of specific training. Moreover, we remark that the performance of the Transformer may be further improved by training (or fine-tuning) it on noisy data, thus aligning the training and testing conditions.}

\begin{table*} 
    \centering
    \caption{Simulation: settings and outcomes. The  reported  $\rm {rmse}$ refers to multi-step ahead simulation. The training time for the WH class includes the time to train the LTI model class used to initialize the optimization. }
    \label{tab:sim}
    \resizebox{0.99\textwidth}{!}{%
    \begin{tabular}{lccccccccc}
        \toprule
        $p(\D)$ & $n_{\text{param}}$ & $n_{\text{layers}}$ & $n_{\text{heads}}$ & $d_{\text{model}}$ & $n_{\text{ctx,enc/dec}}$ & $n_{\rm itr}$ & batch size $b$ & train time & rmse \\
        \midrule
        LTI & 5.6~M & 12 & 4 & 128 & 400/100 & 1M & 32 & 0.9~d & 0.013 \\
        WH & 5.6~M  & 12 & 4 & 128 & 400/100 & 5M & 32 & (0.9 + 4.5)~d& 0.112 \\
        \bottomrule
    \end{tabular}%
    }
\end{table*}

\begin{table}
    \centering
    \caption{Simulation: average $\rm rmse$ obtained by the Transformer and by traditional system identification techniques (subspace and NLSQ for the LTI and WH classes, respectively) for increasing levels of the noise standard deviation $\sigma_e$ corrupting the context output 
    $y_{1:m}$.  {The reported $\rm rmse$ is computed with respect to a  noise-free output sequence $y_{m+1:\nsamp}$}}.
    \label{tab:noise_effect}
    \begin{tabular}{c | l | l | l | l}
    \toprule
        $\sigma_e$ & \multicolumn{2}{c|}{LTI} & \multicolumn{2}{c}{WH} \\
        & Transformer & Subspace & Transformer & NLSQ \\
        \midrule
        0.0 & 0.013 & 0.007 & 0.112 & 0.072 \\
        0.1 & 0.049 & 0.092 & 0.172 & 0.101 \\
        0.2 & 0.100 & 0.113 & 0.248 & 0.136 \\
        0.3 & 0.148 & 0.130 & 0.363 & 0.207 \\
        0.4 & 0.197 & 0.148 & 0.533 & 0.302 \\
        0.5 & 0.249 & 0.165 & 0.703 & 0.378 \\
        \bottomrule
    \end{tabular}
\end{table}

\section{Conclusions and future works}
The novel in-context learning paradigm for model-free system identification presented in this paper enables inferring the behavior of dynamical systems by observing the behavior of other systems within the same class. Compared to traditional system identification, our approach provides a training-free, few-example learning framework. This not only reduces the computational cost when adapting to new tasks, but also obviates the need for specific selections of the dynamical model structure, learning algorithm, fine-tuning of hyper-parameters, etc. 

The proposed work paves the way for new research directions and raises questions that are currently being addressed by the authors and may also intrigue other researchers. The potential areas of investigation include, but are not limited to:

\begin{itemize}

    \item \textbf{From class-to-class and from class-to-system learning:} Transfer learning techniques may be applied. For instance,  the weights of a meta-model pretrained on a specific class of systems (e.g., robotic manipulators) can be adapted to describe another class (e.g., soft robots). Similarly, the meta-model pretrained on a model class can be fine tuned (or distilled) to a specific system instance. 

    \item \textbf{Role of Noise in Learning:} It remains ambiguous whether considering noisy output data in the context during training is beneficial or counterproductive. Standard system identification suggests that noise   degrades the quality of estimates. However, considering that real-world applications of this approach will involve a noisy output history, introducing noise during training might assist in learning how to filter out the noise and make accurate predictions based on a noisy context. A comprehensive statistical analysis is essential to answer this research question.

    \item \textbf{Curriculum Learning:} According to a curriculum learning strategy \cite{wang2021survey},  the Transformer's weights can be optimized using a few-step ahead strategy. The more computationally-intensive simulation criterion can then be employed at a later stage to re-estimate only a subset of the weights. Indeed, the former  is less computationally demanding, making it more suited for massive  learning.  Similarly, as already discussed in the example on model-free simulation (Section \ref{sec:exp_sim}), we can learn a meta-model on a narrow model class (e.g., LTI systems) and then extend the training to a richer one.

    \item \textbf{Dynamical System Generation:} Currently, instances of dynamical systems and their corresponding input/output trajectories are drawn from an arbitrarily chosen prior distribution. There is room for the development of strategies to produce systems and signals that maximize their   informativeness across successive training iterations.
    
    \item \textbf{Data Augmentation in Classic Parametric System Identification:} The proposed method can also serve as a data-augmentation technique. Here, synthetic data generated by the Transformer, based on a context generated by the system under study, can serve as additional input-output data to estimate a parsimonious parametric model of the system.
\end{itemize}

\section*{Acknowledgements}
This work was supported in part by the European Union, Grant Agreement n. 101095672 (project PRAESIIDIUM “Physics informed machine learning-based prediction and reversion of impaired fasting glucose management”, call HORIZON-HLTH-2022- STAYHLTH-02) 
and by the Swiss State Secretariat for Education, Research and lnnovation (SERI) under contract number 22.00463.
The activities of D. Piga have been partly supported  by HASLER STIFTUNG under the project \emph{LEWIS: Learning from wise: a new machine learning paradigm to leverage model rationality.}

\bibliographystyle{plain}
\bibliography{biblio}

\end{document}